\documentclass[conference]{IEEEtran}
\usepackage{amssymb,amsmath,epsfig,graphicx,theorem}
\usepackage{amsfonts}
\usepackage{bm}
\newtheorem{Theo}{Theorem}
\newtheorem{Lem}{Lemma}

\begin{document}
\IEEEoverridecommandlockouts
\title{A Single-letter Upper Bound for the Sum Rate of Multiple Access Channels with Correlated Sources}

\author{\authorblockN{Wei Kang \qquad \qquad Sennur Ulukus}
\authorblockA{Department of Electrical and Computer Engineering\\
University of Maryland,
College Park, MD  20742\\
\emph{wkang@eng.umd.edu \qquad \qquad ulukus@umd.edu}} 
\thanks{This work was supported by NSF Grants CCR $03$-$11311$, CCF $04$-$47613$ and CCF $05$-$14846$;
and ARL/CTA Grant DAAD $19$-$01$-$2$-$0011$.}}


\maketitle

\begin{abstract}
The capacity region of the multiple access channel with arbitrarily correlated sources
remains  an open problem.
Cover, El Gamal and Salehi gave an achievable region in the form of single-letter
entropy and mutual information  expressions, without a single-letter converse.
Cover, El Gamal and Salehi also gave a converse in terms of some $n$-letter mutual informations,
which are incomputable.
In this paper, we derive an upper bound for the sum rate of this channel in
a single-letter expression by using spectrum analysis.
The incomputability of the sum rate of Cover, El Gamal and Salehi scheme
comes from the difficulty of characterizing the possible joint distributions for
the $n$-letter channel inputs. Here we introduce a new data processing inequality,
which leads to a single-letter necessary condition for these possible joint distributions.
We develop a single-letter upper bound for the sum rate by using this single-letter necessary
condition on the possible joint distributions.
\end{abstract}

\section{Introduction}\label{intro}
The problem of determining the capacity region of the multiple access channel
with correlated sources can be formulated as follows.
Given a pair of correlated sources $(U,V)$ described by the joint probability distribution $p(u,v)$, and
a discrete, memoryless, multiple access channel characterized by
the transition probability
$p(y|x_1, x_2)$, what are the necessary and sufficient conditions for the
reliable transmission of $n$ independent identically distributed (i.i.d.)
samples of the sources through the channel, in $n$ channel uses, as $n\rightarrow\infty$?

This problem was studied by Cover, El Gamal and Salehi in \cite{Cover:1980},
where an achievable region expressed by single-letter entropies and
 mutual informations was given. This region was shown to be suboptimal
by Dueck \cite{Dueck:1981}.
Cover, El Gamal and Salehi \cite{Cover:1980} also provided a capacity result with both
achievability and converse
in  incomputable expressions in the form of some
$n$-letter mutual informations.
In this paper, we derive an upper bound for the sum rate of this channel
in a single-letter expression. 

The
incomputability of the sum rate of Cover, El Gamal and Salehi scheme is due to the difficulty
of characterizing the possible joint distributions for the $n$-letter channel inputs.
The Cover, El Gamal, Salehi converse is 
\begin{equation}\label{cgs}
H(U,V)\le \frac{1}{n} I(X_1^n, X_2^n;Y^n)
\end{equation}
where the random variables involved have a joint distribution expressed in the form
\begin{equation}\label{cgsmc}
\prod_{i=1}^np(u_i, v_i)p(x_1^n|u^n)p(x_2^n|v^n) \prod_{i=1}^np(y_i|x_{1i}, x_{2i})
\end{equation}
i.e., the sources and the channel inputs satisfy the Markov chain relation
$X_1^n\rightarrow U^n\rightarrow V^n\rightarrow X_2^n$. 
It is difficult to evaluate the mutual information on the right hand side of (\ref{cgs})
when the joint probability distribution of the random variables involved is subject to
(\ref{cgsmc}).

A usual way to upper bound the mutual information in (\ref{cgs}) is
\begin{align}
\frac{1}{n}I(X_1^n, X_2^n;Y^n)&\le\frac{1}{n}\sum_{i=1}^n I(X_{1i}, X_{2i};Y_i)\nonumber\\
&\le\max I(X_1, X_2;Y)\label{fib}
\end{align}
where the maximization in (\ref{fib}) is over all possible $X_1$ and $X_2$ such that
$X_1\rightarrow U^n\rightarrow V^n\rightarrow X_2$.
Therefore, combining (\ref{cgs}) and (\ref{fib}), a single-letter upper bound for the sum rate is 
obtained as,
\begin{equation}H(U,V)\le \max I(X_1, X_2;Y)\label{target}
\end{equation}
where the maximization is over all $X_1, X_2$ such that $X_1\rightarrow U^n\rightarrow V^n\rightarrow X_2$.
%
However, a closed form expression
for $p(x_1, x_2)$ satisfying this Markov chain, for all $U$, $V$ and $n$, seems intractable
to obtain.

Data processing inequality \cite[p. 32]{Cover:1991}
is an intuitive way to obtain a necessary condition on $p(x_1, x_2)$ for
 the above Markov chain constraint, i.e., we may try to solve the following problem
 as an upper bound for (\ref{target})
 \begin{align}
\max\quad &I(X_1, X_2;Y)\\
 \text{ s.t.}\quad &I(X_1;X_2)\le I(U^n;V^n)=nI(U,V)\nonumber
 \end{align}
 where ``s.t." line provides a constraint on the feasible set of $p(x_1, x_2)$.
However, when $n$ is large, this upper bound becomes trivial as
$nI(U,V)$ quickly gets larger than  $I(X_1;X_2)$ for $p(x_1, x_2)$ even without the
Markov chain constraint. Although the data processing inequality in its usual form
does not prove useful in this problem, we will still use the basic methodology of employing
a data processing inequality to represent the Markov chain constraint on the valid input
distributions. For this, we will introduce a new data processing inequality.

Spectrum analysis has been instrumental in the study of some properties of pairs of correlated random variables,
especially, those of the i.i.d. sequences of pairs of correlated random variables,
e.g.,~common information in \cite{Witsenhausen:1975}
and isomorphism in \cite{Marton:1981}.
In this paper, we use spectrum analysis to introduce a new data processing inequality.
Our new data processing inequality provides a single-letter necessary condition for the joint
distributions satisfying the Markov chain condition, and leads to a non-trivial single-letter upper bound 
for the sum rate of the multiple access channel 
with correlated sources.

\section{Some Preliminaries}
In this section, we provide some basic results what will be used in our later development.
The concepts used here are originally introduced by Witsenhausen in \cite{Witsenhausen:1975}
 in the context of operator theory.
Here,  we limit ourselves to the finite alphabet case,  and  derive our results by means of matrix theory.

We first introduce our matrix notation for probability distributions.
For a pair of discrete random variables $X$ and $Y$,
which take values in $\mathcal{X}=\{x_1, x_2,\dots, x_m\}$
and $\mathcal{Y}=\{y_1, y_2,\dots, y_n\}$, respectively,
the joint distribution matrix $P_{XY}$ is defined as
$P_{XY}(i,j)\triangleq Pr(X=x_i, Y=y_j)$,
where $P_{XY}(i,j)$ denotes the $(i,j)$-th element of the matrix $P_{XY}$.
From this definition, we have $P_{XY}^T=P_{YX}$.
The marginal distribution of a random variable $X$ is defined as
a diagonal matrix with $P_{X}(i,i)\triangleq Pr(X=x_i)$.
The vector-form marginal distribution is defined as $p_X(i)\triangleq Pr(X=x_i)$,
i.e.,~$p_X=P_X \mathbf{e}$, where $\mathbf{e}$ is a vector of all ones.
Similarly, we define $p_X^{\frac{1}{2}}\triangleq P_X^{\frac{1}{2}} \mathbf{e}$ 
and $p_X^{-\frac{1}{2}}\triangleq P_X^{-\frac{1}{2}} \mathbf{e}$.
The conditional distribution of $X$ given $Y$ is defined in the
matrix form as $P_{X|Y}(i,j)\triangleq Pr(X=x_i|Y=y_j)$, and
$P_{X|Y}=P_{XY}P_{Y}^{-1}$.

We define a new quantity, $\tilde{P}_{XY}$, which will
play an important role in the rest of the paper, as
\begin{equation}
\tilde{P}_{XY}=P_X^{-\frac{1}{2}}P_{XY}P_Y^{-\frac{1}{2}}\label{def}
\end{equation}

Our main theorem in this section identifies the spectral properties of $\tilde{P}_{XY}$.
Before stating our theorem, we provide the following lemma, which will be used
in its proof.

\begin{Lem}\cite[p. 49]{Berman:1979}\label{sto}
The spectral radius of a stochastic matrix is $1$.
A non-negative matrix $T$ is stochastic if and only if $\mathbf{e}$ is
an eigenvector of $T$ corresponding to the eigenvalue $1$.
\end{Lem}

\begin{Theo}\label{iff}
An $m\times n$ non-negative matrix $P$ is a joint distribution matrix
with marginal distributions $P_X$ and $P_Y$, i.e.,~$P\mathbf{e}=p_X\triangleq P_X\mathbf{e}$ and
$P^T\mathbf{e}=p_Y\triangleq P_Y\mathbf{e}$,
if and only if the singular value decomposition (SVD) of $\tilde{P}\triangleq P_X^{-\frac{1}{2}}PP_Y^{-\frac{1}{2}}$ 
satisfies
\begin{equation}
\tilde{P}=U\Lambda V^T =
p_X^{\frac{1}{2}}(p_Y^{\frac{1}{2}})^T+\sum_{i=2}^l \lambda_i \mathbf{u}_i\mathbf{v}_i^T\label{fun}
\end{equation}
where $U\triangleq [\mathbf{u}_1,\dots, \mathbf{u}_l]$ and $V\triangleq [\mathbf{v}_1,\dots, \mathbf{v}_l]$
are two unitary matrices, $\Lambda\triangleq \mathrm{diag}[\lambda_1,\dots,\lambda_l]$ and $l=\min(m,n)$;
$\mathbf{u}_1=p_X^{\frac{1}{2}}$, $\mathbf{v}_1=p_Y^{\frac{1}{2}}$, and
$\lambda_1=1\ge\lambda_2\ge\dots\ge\lambda_l\ge 0$. That is, all of the singular values of $\tilde{P}$ are between $0$ and 
$1$, the largest singular value of $\tilde{P}$ is $1$, and the corresponding left and right singular vectors are $p_X^{\frac{1}{2}}$
and $p_Y^{\frac{1}{2}}$.
\end{Theo}
\begin{proof} Let $\tilde{P}$ satisfy (\ref{fun}),
then
\begin{align}
P_X^{\frac{1}{2}}\tilde{P} P_Y^{\frac{1}{2}}\mathbf{e}&=P_X^{\frac{1}{2}}
\left(p_X^{\frac{1}{2}}(p_Y^{\frac{1}{2}})^T+\sum_{i=2}^l \lambda_i \mathbf{u}_i\mathbf{v}_i^T\right)
p_Y^{\frac{1}{2}}\nonumber\\
&=P_X^{\frac{1}{2}}p_X^{\frac{1}{2}}(p_Y^{\frac{1}{2}})^T p_Y^{\frac{1}{2}}+
P_X^{\frac{1}{2}}\sum_{i=2}^l \lambda_i \mathbf{u}_i\mathbf{v}_i^T \mathbf{v}_1\nonumber\\
&=p_X
\end{align}
Similarly, $\mathbf{e}^T P_X^{\frac{1}{2}}\tilde{P} P_Y^{\frac{1}{2}}=p_Y^T$.
Thus, the non-negative matrix $P_X^{\frac{1}{2}}\tilde{P} P_Y^{\frac{1}{2}}$ is a joint distribution matrix with
marginal distributions $p_X$ and $p_Y$.

Conversely, we consider a joint distribution $P$ with marginal distributions $p_X$ and $p_Y$.
We need to show that the singular values of $\tilde{P}$ lie in $[0,1]$,
the largest singular value is  equal to $1$, and
$p_X^{\frac{1}{2}}$ and $p_Y^{\frac{1}{2}}$, respectively, are
the left and right singular vectors
corresponding to the singular value $1$.

To this end, we first construct a Markov chain $X\rightarrow Y\rightarrow Z$ with $P_{XY}=P_{ZY}=P$.
Note that this also implies $P_X=P_Z$, $\tilde{P}_{XY}=\tilde{P}_{ZY}=\tilde{P}$,  and $P_{X|Y}=P_{Z|Y}$.
The special structure of the constructed Markov chain provides the following:
\begin{align}
P_{X|Z}&=P_{X|Y}P_{Y|Z}=P_{X|Y}P_{Y|X}=PP_Y^{-1}P^T P_X^{-1}\nonumber\\
&=P_X^{\frac{1}{2}}(P_X^{-\frac{1}{2}}PP_Y^{-\frac{1}{2}})(P_Y^{-\frac{1}{2}}P^T P_X^{-\frac{1}{2}})
P_X^{-\frac{1}{2}}\nonumber\\
&=P_X^{\frac{1}{2}}\tilde{P}\tilde{P}^T P_X^{-\frac{1}{2}}
\end{align}
We note that the matrix $P_{X|Z}$ is similar to the matrix $\tilde{P}\tilde{P}^T$ \cite[p. 44]{Horn:1985}.
Therefore, all eigenvalues of $P_{X|Z}$  are the eigenvalues of $\tilde{P}\tilde{P}^T$ as well, and if 
$\mathbf{v}$ is a left eigenvector of $P_{X|Z}$ corresponding to an eigenvalue $\mu$,  
then $P_X^{\frac{1}{2}}\mathbf{v}$ is a left eigenvector of $\tilde{P}\tilde{P}^T$
corresponding to the same eigenvalue.

We note that $P_{X|Z}$ is a stochastic matrix, 
therefore, from Lemma \ref{sto}, $\mathbf{e}$ is a left eigenvector of $P_{X|Z}$ corresponding
the eigenvalue $1$, which is also equal to the spectral radius of $P_{X|Z}$.
Since $P_{X|Z}$ is similar to $\tilde{P}\tilde{P}^T$, 
we have that $p_{X}^{\frac{1}{2}}$ is a left eigenvector of $\tilde{P}\tilde{P}^T$
with eigenvalue $1$, 
%
and the rest of the eigenvalues of $\tilde{P}\tilde{P}^T$ lie in $[-1,1]$.
In addition, $\tilde{P}\tilde{P}^T$ is a symmetric positive semi-definite matrix,
which implies that the eigenvalues of $\tilde{P}\tilde{P}^T$
are real and non-negative. 
Since the eigenvalues of $\tilde{P}\tilde{P}^T$ 
are non-negative, and the largest eigenvalue is equal to $1$,   we conclude that all of the eigenvalues of $\tilde{P}\tilde{P}^T$
lie in  the interval $[0,1]$.

The singular values of $\tilde{P}$ are the square roots of the eigenvalues
of $\tilde{P}\tilde{P}^T$, and the left singular vectors of $\tilde{P}$ are
 the eigenvectors of $\tilde{P}\tilde{P}^T$.
Thus, the singular values of $\tilde{P}$ lie in $[0,1]$,
the largest singular value is  equal to $1$, and $p_X^{\frac{1}{2}}$
is a left singular vector corresponding to the singular value $1$.
The corresponding right singular vector is
\begin{align}
\mathbf{v}_1^T&=\mathbf{u}_1^T\tilde{P}=(p_X^{\frac{1}{2}})^T P_X^{-\frac{1}{2}}P P_Y^{-\frac{1}{2}}
=p_Y^T P_Y^{-\frac{1}{2}}
=(p_Y^{\frac{1}{2}})^T
\end{align}
which concludes the proof.
\end{proof}
\section{A New Data Processing Inequality}\label{dpi}
In this section, we introduce a new data processing inequality in the following theorem.
We first provide a lemma that will be used in its proof.
\begin{Lem}\label{pro}\cite[p. 178]{Horn:1991}
For matrices $A$ and $B$
\begin{equation}
\lambda_{i}(AB)\le\lambda_{i}(A)\lambda_1(B)
\end{equation}
where $\lambda_{i}(\cdot)$ denotes the $i$-th largest singular value of a matrix.
\end{Lem}
\begin{Theo}\label{sigpro}
If $X\rightarrow Y\rightarrow Z$, then
\begin{align}
\lambda_i(\tilde{P}_{XZ})\le\lambda_i(\tilde{P}_{XY})\lambda_2(\tilde{P}_{YZ})&\le\lambda_i(\tilde{P}_{XY})
 \end{align}
where  $i=2,\dots, \mathrm{rank}(\tilde{P}_{XZ})$.
\end{Theo}
\begin{proof}
From the structure of the Markov chain, and from the definition of $\tilde{P}_{XY}$ in (\ref{def}), we have
\begin{align}
\tilde{P}_{XZ}&=P_X^{-\frac{1}{2}}P_{XZ}P_Z^{-\frac{1}{2}}
=\tilde{P}_{XY}\tilde{P}_{YZ}\label{prod}
\end{align}
Using (\ref{fun}) for $\tilde{P}_{XZ}$, we obtain
\begin{align}
\tilde{P}_{XZ}=&p_X^{\frac{1}{2}}(p_Z^{\frac{1}{2}})^T+
\sum_{i=2}^l \lambda_i(\tilde{P}_{XZ}) \mathbf{u}_i(\tilde{P}_{XZ})\mathbf{v}_i(\tilde{P}_{XZ})^T\label{lef}
\end{align}
and using (\ref{fun}) for $\tilde{P}_{XY}$ and $\tilde{P}_{YZ}$ yields
\begin{align}
\tilde{P}_{XY}\tilde{P}_{YZ}
=&\left(p_X^{\frac{1}{2}}(p_Y^{\frac{1}{2}})^T\!+\!
\sum_{i=2}^l \lambda_i(\tilde{P}_{XY}) \mathbf{u}_i(\tilde{P}_{XY})\mathbf{v}_i(\tilde{P}_{XY})^T\right)\nonumber\\
\times &\left(p_Y^{\frac{1}{2}}(p_Z^{\frac{1}{2}})^T+
\sum_{i=2}^l \lambda_i(\tilde{P}_{YZ}) \mathbf{u}_i(\tilde{P}_{YZ})\mathbf{v}_i(\tilde{P}_{YZ})^T\right)\nonumber\\
=&p_X^{\frac{1}{2}}(p_Z^{\frac{1}{2}})^T+
\left(\sum_{i=2}^l \lambda_i(\tilde{P}_{XY}) \mathbf{u}_i(\tilde{P}_{XY})\mathbf{v}_i(\tilde{P}_{XY})^T\right)
\nonumber\\
\times&\left(\sum_{i=2}^l \lambda_i(\tilde{P}_{YZ}) \mathbf{u}_i(\tilde{P}_{YZ})\mathbf{v}_i(\tilde{P}_{YZ})^T\right)
\label{righ}
\end{align}
where the two cross-terms vanish since $p_Y^{\frac{1}{2}}$ is
both $\mathbf{v}_1(\tilde{P}_{XY})$ and $\mathbf{u}_1(\tilde{P}_{YZ})$, and therefore, $p_Y^{\frac{1}{2}}$ is
orthogonal to both $\mathbf{v}_i(\tilde{P}_{XY})$ and $\mathbf{u}_j(\tilde{P}_{YZ})$, for all $i,j \neq 1$.
Using (\ref{prod}) and equating (\ref{lef}) and (\ref{righ}), we obtain
\begin{align}
\sum_{i=2}^l \lambda_i(\tilde{P}_{XZ}) &\mathbf{u}_i(\tilde{P}_{XZ})\mathbf{v}_i(\tilde{P}_{XZ})^T\nonumber\\
=&\left(\sum_{i=2}^l \lambda_i(\tilde{P}_{XY}) \mathbf{u}_i(\tilde{P}_{XY})\mathbf{v}_i(\tilde{P}_{XY})^T\right)
\nonumber\\
&\times \left(\sum_{i=2}^l \lambda_i(\tilde{P}_{YZ}) \mathbf{u}_i(\tilde{P}_{YZ})\mathbf{v}_i(\tilde{P}_{YZ})^T\right)
\label{bela}
\end{align}
The proof is completed by applying Lemma \ref{pro} to (\ref{bela}).
\end{proof}
\section{On i.i.d. Sequences}\label{iid}
Let $(X^n, Y^n)$ be a pair of i.i.d. sequences,
where each pair of letters of these sequences satisfies a joint distribution $P_{XY}$.
Thus, the joint distribution of the sequences is $P_{X^n Y^n}=P_{XY}^{\otimes n}$,
where $A^{\otimes1}\triangleq A$, $A^{\otimes k}\triangleq A\otimes A^{\otimes (k-1)}$,
and $\otimes$ represents the Kronecker product of matrices \cite{Horn:1985}.

From (\ref{def}),
\begin{equation}
P_{XY}=P_X^{\frac{1}{2}}\tilde{P}_{XY}P_Y^{\frac{1}{2}}
\end{equation}
Then,
\begin{equation}
P_{X^n Y^n}=P_{XY}^{\otimes n}=(P_X^{\frac{1}{2}}\tilde{P}_{XY}P_Y^{\frac{1}{2}})^{\otimes n}
=(P_X^{\frac{1}{2}})^{\otimes n}\tilde{P}_{XY}^{\otimes n}(P_Y^{\frac{1}{2}})^{\otimes n}
\end{equation}
We also have $P_{X_1^n}=(P_X)^{\otimes n}$ and $P_{Y_1^n}=(P_Y)^{\otimes n}$. Thus,
\begin{align}
\tilde{P}_{X^nY^n}&\triangleq P_{X^n}^{-\frac{1}{2}} P_{X^n Y^n} P_{Y^n}^{-\frac{1}{2}}\nonumber\\
&=(P_X^{-\frac{1}{2}})^{\otimes n}(P_X^{\frac{1}{2}})^{\otimes n}\tilde{P}_{XY}^{\otimes n}
(P_Y^{\frac{1}{2}})^{\otimes n}(P_Y^{-\frac{1}{2}})^{\otimes n}\nonumber\\
&=\tilde{P}_{XY}^{\otimes n}
\end{align}

Applying SVD to $\tilde{P}_{X^nY^n}$, we have
\begin{equation}
\tilde{P}_{X^nY^n}=U_n\Lambda_nV_n^T=\tilde{P}_{XY}^{\otimes n}=U^{\otimes n}\Lambda^{\otimes n}(V^{\otimes n})^T
\end{equation}
From the uniqueness of the SVD, we know that $U_n=U^{\otimes n}$, $\Lambda_n=\Lambda^{\otimes n}$
and $V_n=V^{\otimes n}$. Then, the ordered singular values of $\tilde{P}_{X^nY^n}$ are
\begin{equation}
\{1, \lambda_2(\tilde{P}_{XY}), \dots, \lambda_2(\tilde{P}_{XY}),\dots\}\nonumber
\end{equation}
where the second through the $n+1$-st singular values are all equal to $\lambda_2(\tilde{P}_{XY})$.
\section{A Necessary Condition}
As stated in Section \ref{intro}, the sum rate can be upper bounded as
\begin{equation}
H(U,V)\le \max I(X_1, X_2;Y)
\end{equation}
where the maximization  is over all possible $X_1$ and $X_2$ that satisfy the Markov chain
$X_1\rightarrow U^n\rightarrow V^n\rightarrow X_2$.
%

From Theorem \ref{sigpro} in Section \ref{dpi}, we know that if $X_1\rightarrow U^n\rightarrow V^n\rightarrow X_2$,
then, for $i=2,\dots, \text{rank}(\tilde{P}_{X_1X_2})$,
\begin{equation}
\lambda_i(\tilde{P}_{X_1X_2})\le\lambda_2(\tilde{P}_{X_1U^n})\lambda_i(\tilde{P}_{U^nV^n})\lambda_2(\tilde{P}_{V^nX_2})
\end{equation}
We showed in Section \ref{iid} that $\lambda_i(\tilde{P}_{U^nV^n})\le\lambda_2(\tilde{P}_{UV})$ for $i\ge 2$, and
 $\lambda_i(\tilde{P}_{U^nV^n})=\lambda_2(\tilde{P}_{UV})$ for $i=2,\dots,n+1$.
Therefore,   for $i=2,\dots, \text{rank}(\tilde{P}_{X_1X_2})$, we have
\begin{equation}\label{ine}
\lambda_i(\tilde{P}_{X_1X_2})\le\lambda_2(\tilde{P}_{X_1U^n})\lambda_2(\tilde{P}_{UV})\lambda_2(\tilde{P}_{V^nX_2})
\end{equation}
From Theorem \ref{iff}, we know that $\lambda_2(\tilde{P}_{X_1U^n})\le 1$ and $\lambda_2(\tilde{P}_{V^nX_2})\le 1$.
Next, in  Theorem \ref{app},  we determine that the least upper bound for
$\lambda_2(\tilde{P}_{X_1U^n})$ and $\lambda_2(\tilde{P}_{V^nX_2})$ is also $1$.
\begin{Theo}\label{app}
Let $F(n, P_{X_1})$ be the set of all  joint distributions for $X_1$ and $U^n$
with a given marginal distribution for $X_1$, $P_{X_1}$. Then,
\begin{equation}
\sup_{F(n, P_{X_1}),\; n=1,2,\dots}\lambda_2(\tilde{P}_{X_1U^n})=1\label{mapping}
\end{equation}
\end{Theo}
The proof of Theorem  \ref{app} is given in the Appendix.

Combining (\ref{ine}) and Theorem \ref{app},  we obtain the main result of our paper, 
which is stated in the following theorem.
\begin{Theo}
If a pair of i.i.d. sources $(U,V)$ with joint distribution $P_{UV}$ can be transmitted reliably 
through a discrete, memoryless, multiple access channel characterized by $P_{Y|X_1X_2}$,
then
\begin{equation}
H(U,V)\le I(X_1,X_2;Y)
\end{equation}
for some $(X_1, X_2)$ with
\begin{equation}\lambda_{i}(\tilde{P}_{X_1X_2})\le\lambda_2(\tilde{P}_{UV}),\quad
i=2,\dots, \mathrm{rank}(\tilde{P}_{X_1X_2}).
\end{equation}
\end{Theo}
\section{Some Simple Examples}
We consider a multiple access channel where the alphabets of
$X_1$, $X_2$ and $Y$ are all binary, and the channel transition probability matrix
$p(y|x_1, x_2)$ is given as
\begin{equation}
\begin{array}{c|cccc}
Y\backslash X_1X_2&11&10&01&00\\
\hline
1&1&1/2&1/2&0\\
0&0&1/2&1/2&1
\end{array}\nonumber
\end{equation}
The following is a trivial upper bound, which we provide 
 as a benchmark,
\begin{equation}\label{triv}
\underset{p(x_1, x_2)}{\max} I(X_1, X_2;Y)=1
\end{equation}
where the maximization is over all binary bivariate distributions.
The maximum is achieved by
 $P(X_1=1,X_2=1)=P(X_1=0, X_2=0)=1/2$.
 We note that this upper bound does not depend on the source distribution.

First, we consider a binary source $(U,V)$ with the following joint distribution
$p(u,v)$
\begin{equation}
\begin{array}{c|cc}
U\backslash V&1&0\\
\hline
1&1/3&1/6\\
0&1/6&1/3
\end{array}\nonumber
\end{equation}
In this case, $H(U,V)=1.92$.
We first note, using the trivial upper bound in (\ref{triv}), that,
it is impossible to transmit this source through the given channel reliably.
The upper bound we developed in this paper gives $2/3$ for this source. We also note that,
for this case, our upper bound coincides with the single-letter achievability
expression given in \cite{Cover:1980}, which is
\begin{equation}\label{cgsa}
H(U,V)\le I(X_1, X_2;Y)
\end{equation}
where $X_1, X_2$ are such that $X_1\rightarrow U\rightarrow V\rightarrow X_2$ holds. 
Therefore, for this case, our upper bound is the converse, as it matches the achievability expression.

Next, we consider a binary source $(U,V)$ with the following joint distribution $p(u,v)$
\begin{equation}
\begin{array}{c|cc}
U\backslash V&1&0\\
\hline
1&0&0.1\\
0&0.1&0.8
\end{array}\nonumber
\end{equation}
In this case, $H(U,V)=0.92$, the single-letter achievability in (\ref{cgsa}) reaches $0.51$
and our upper bound is $0.56$. The gap between the achievability and our upper bound is quite small.
We note that, in this case, the trivial upper bound in (\ref{triv})  fails to test
whether it is
possible to have reliable transmission or not, while our upper bound determines conclusively that reliable
transmission is not possible.

Finally, we consider  a binary source $(U,V)$ with the following joint distribution $p(u,v)$
\begin{equation}
\begin{array}{c|cc}
U\backslash V&1&0\\
\hline
1&0&0.85\\
0&0.1&0.05
\end{array}\nonumber
\end{equation}
In this case, $H(U,V)=0.75$, the single-letter achievability expression in (\ref{cgsa}) gives $0.57$ 
and our upper bound is $0.9$. We note that the joint entropy of the sources falls into the gap
between the achievability expression and our upper bound, which means that we cannot
conclude whether it is possible (or not) to transmit these sources  through the 
channel reliably.

\section{Conclusion}
In this paper, we investigated the problem of transmitting correlated sources through a multiple access channel.
We utilized the spectrum analysis to develop a new data processing
inequality, which provided a single-letter necessary condition for the joint distributions
satisfying the Markov chain condition.  
By using our new data processing inequality, we developed a new single-letter upper bound
for the sum rate of the multiple access channel with correlated sources.

\appendix[Proof of Theorem \ref{app}]
To find $\underset{F(n, P_{X_1}),\; n=1,2,\dots}{\sup} \lambda_2(\tilde{P}_{X_1U^n})$,
we need to exhaust the sets $F(n, P_{X_1})$ with $n\ge 1$.
In the following, we show that it suffices to check only the asymptotic case. 

For any joint distribution $P_{X_1U^n}\in F(n, P_{X_1})$,
we attach an independent $U$, say $U_{n+1}$, to the existing $n$-sequence,
and get a new joint distribution $P_{X_1U^{n+1}}=P_{X_1U^n}\otimes p_{U}$,
where $p_{U}$ is the marginal distribution of $U$ in the vector form.
By arguments similar to those in Section \ref{iid},
we have that
$\lambda_i(\tilde{P}_{X_1U^{n+1}})=\lambda_i(\tilde{P}_{X_1U^{n}})$.
Therefore, for every $P_{X_1U^n}\in F(n, P_{X_1})$, there exists some $P_{X_1U^{n+1}}\in F(n+1, P_{X_1})$, 
such that $\lambda_i(\tilde{P}_{X_1U^{n+1}})=\lambda_i(\tilde{P}_{X_1U^{n}})$.
Thus,
\begin{equation}
\underset{F(n, P_{X_1})}{\sup} \lambda_2(\tilde{P}_{X_1U^n})\le
\underset{F(n+1, P_{X_1})}{\sup} \lambda_2(\tilde{P}_{X_1U^{n+1}})\label{mono}
\end{equation}
From (\ref{mono}),  we see that $\underset{F(n, P_{X_1})}{\sup} \lambda_2(\tilde{P}_{X_1U^n})$
is monotonically non-decreasing in $n$. We also note that $\lambda_{2}(\tilde{P}_{X_1U^n})$ is upper bounded by $1$
for all $n$, i.e.,~$\lambda_{2}(\tilde{P}_{X_1U^n})\le 1$.
 Therefore,
\begin{equation}
\underset{F(n, P_{X_1}),\; n=1,2,\dots}{\sup} \lambda_2(\tilde{P}_{X_1U^n})=
\lim_{n\rightarrow\infty}\underset{F(n, P_{X_1})}{\sup} \lambda_2(\tilde{P}_{X_1U^n})\label{lim}
\end{equation}
To complete the proof, we need the following lemma. 
\begin{Lem}\label{decom}\cite{Witsenhausen:1975}
$\lambda_{2}(\tilde{P}_{XY})=1$ if and only if $P_{XY}$ decomposes.
By $P_{XY}$ decomposes, we mean that there exist sets $S_1\in\mathcal{X}$,
$S_2\in\mathcal{Y}$, such that $P(S_1)$, $P(\mathcal{X}-S_1)$, $P(S_2)$, $P(\mathcal{Y}-S_2)$
are positive, while $P((\mathcal{X}-S_1)\times S_2)=P(S_1\times (\mathcal{Y}-S_2))=0$.
\end{Lem}

In the following, we will show by construction that 
there exists a joint distribution that 
decomposes asymptotically.

For a given marginal distribution $P_{X_1}$, we arbitrarily choose a subset $S_1$ from
 the alphabet of $X_1$. We  find a set $S_2$ in the alphabet of $U^n$ such that
 $P(S_1)=P(S_2)$ if it is possible. Otherwise, we pick $S_2$ such that $|P(S_1)-P(S_2)|$ is minimized.
We denote $\mathcal{S}(n)$ to be the set of all subsets of the alphabet
 of $U^n$ and we also define $P_{\max}=\max Pr(s)$ for all $s\in \mathcal{U}$.
 Then, we have
\begin{equation}\label{upb}
\underset{S_2\subset\mathcal{S}(n)}{\min}|P(S_2)-P(S_1)|\le P_{\max}^n
\end{equation}

We construct a joint distribution for $X_1$ and $U^n$ as follows.
First, we construct the joint distribution $P^i$ corresponding to the case where 
$X_1$ and $U^n$ are independent.
Second, we rearrange the alphabets of $X_1$ and $U^n$ and group the sets
 $S_1$, $\mathcal{X}_1-S_1$, $S_2$ and $\mathcal{U}^n-S_2$ as follows
\begin{equation}
P^i=\left[\begin{array}{ll}P_{11}^i&P_{12}^i\\P_{21}^i&P_{22}^i\end{array}\right]\label{Pi}
\end{equation}
where $P_{11}^i$, $P_{12}^i$, $P_{21}^i$, $P_{22}^i$    correspond to the sets $S_1\times S_2$,
$S_1\times (\mathcal{U}^n-S_2)$,
$(\mathcal{X}^1-S_1)\times S_2$,
$(\mathcal{X}^1-S_1)\times (\mathcal{U}^n-S_2)$, respectively.
Here, we assume that $P(S_2)\ge P(S_1)$.
Then, we scale these four sub-matrices as
$P_{11}=\frac{P_{11}^iP(S_1)}{P(S_1)P(S_2)}$,
$P_{12}=0$,
$P_{21}=\frac{P_{21}^i(P(S_2)-P(S_1))}{(1-P(S_1))P(S_2)}$,
$P_{22}=\frac{P_{21}^i(1-P(S_2))}{(1-P(S_1))(1-P(S_2))}$,
and let
\begin{equation}
P=\left[\begin{array}{ll}P_{11}&0\\P_{21}&P_{22}\end{array}\right]
\end{equation}
We note that $P$ is a joint distribution for $X_1$ and $U^n$ with the given marginal
distributions.
Next, we move the mass in the sub-matrix $P_{21}$ to $P_{11}$, which yields
\begin{equation}
P'\!\triangleq\!\left[\begin{array}{ll}P_{11}'&0\\0&P_{22}\end{array}\right]
\!\!=P+E=\!\!\left[\begin{array}{ll}P_{11}&0\\P_{21}&P_{22}\end{array}\right]
+\left[\begin{array}{ll}E_{11}&0\\-E_{21}&0\end{array}\right]\label{Pp}
\end{equation}
where
$E_{21}\triangleq P_{21}$, $E_{11}\triangleq\frac{P_{11}^i(P(S_2)-P(S_1))}{P(S_1)P(S_2)}$,
and $P_{11}'=\frac{P_{11}P(S_2)}{P(S_1)}$.
We denote $P'_{X_1}$ and $P'_{U^n}$ as the marginal distributions of $P'$.
We note that $P'_{U^n}=P_{U^n}$ and $P'_{X_1}=P_{X_1}M$ where $M$ is a scaling
diagonal matrix. The elements in the set $S_1$ are scaled up by a factor of $\frac{P(S_2)}{P(S_1)}$,
and those in the set $\mathcal{X}_1-S_1$ are scaled down by a factor of $\frac{1-P(S_2)}{1-P(S_1)}$.
Then,
\begin{align}
\tilde{P}'
&=M^{-\frac{1}{2}}\tilde{P}+M^{-\frac{1}{2}}P_{X_1}^{-\frac{1}{2}}EP_{U^n}^{-\frac{1}{2}}
\end{align}
We will need the following lemmas in the remainder of our derivations.
Lemma \ref{times} can be proved using techniques similar to those in the proof of Lemma \ref{sum} \cite{Stewart:1993}.
\begin{Lem}\label{sum}\cite{Stewart:1993}
If $A'=A+E$, then $|\lambda_{i}(A')-\lambda_{i}(A)|\le||E||_2$, where $||E||_2$ is the spectral norm of $E$.
\end{Lem}
\begin{Lem}\label{times}
If $A'=MA$, where $M$ is an invertible matrix,
then $||M^{-1}||_2^{-1}\le\lambda_{i}(A')/\lambda_{i}(A)\le||M||_2$.
\end{Lem}

Since $P'$ decomposes,  using Lemma \ref{decom}, we conclude that  $\lambda_2(\tilde{P}')=1$.
We upper bound  $||P_{X_1}^{-\frac{1}{2}}EP_{U^n}^{-\frac{1}{2}}||_2$ as follows,
\begin{equation}
||P_{X_1}^{-\frac{1}{2}}EP_{U^n}^{-\frac{1}{2}}||_2
\le
||P_{X_1}^{-\frac{1}{2}}EP_{U^n}^{-\frac{1}{2}}||_F
\label{22f}
\end{equation}
where $||\cdot||_F$ is the Frobenius norm.
Combining (\ref{Pi}) and (\ref{Pp}), we have
\begin{align}
||P_{X_1}^{-\frac{1}{2}}&EP_{U^n}^{-\frac{1}{2}}||_F
\le\frac{(P(S_2)-P(S_1))}{P_1'P(S_2)}
||P_{X_1}^{-\frac{1}{2}}P^iP_{U^n}^{-\frac{1}{2}}||_F\label{fup}
\end{align}
where $P_1'\triangleq\min(P(S_1),1-P(S_1))$. Since $P^i$ corresponds to
the independent case, we have $||P_{X_1}^{-\frac{1}{2}}P^iP_{U^n}^{-\frac{1}{2}}||_F=1$
from (\ref{fun}). Then, from (\ref{upb}), (\ref{22f}) and (\ref{fup}),  we obtain
\begin{equation}
||P_{X_1}^{-\frac{1}{2}}EP_{U^n}^{-\frac{1}{2}}||_2\le c_1P_{\max}^n
\end{equation}
where $c_1\triangleq\frac{1}{P_1'P(S_2)}$.

From Lemma \ref{pro}, we have
\begin{align}
||M^{-\frac{1}{2}}&P_{X_1}^{-\frac{1}{2}}EP_{U^n}^{-\frac{1}{2}}||_2=
|\lambda_{1}(M^{-\frac{1}{2}}P_{X_1}^{-\frac{1}{2}}EP_{U^n}^{-\frac{1}{2}})|\nonumber\\
&\le\left(\frac{1-P(S_1)}{1-P(S_2)}\right)^{\frac{1}{2}}c_1P_{\max}^n\triangleq c_2P_{\max}^n
\end{align}
From Lemma \ref{sum}, we have
\begin{equation}
1-c_2P_{\max}^{\frac{n}{2}}\le\lambda_2(M^{-\frac{1}{2}}\tilde{P})\le 1+c_2P_{\max}^{\frac{n}{2}}
\end{equation}
We upper bound $||M^{\frac{1}{2}}||_2$ as follows
\begin{align}
||M^{\frac{1}{2}}||_2=&\sqrt{\frac{P(S_2)}{P(S_1)}}
\le1+\sqrt{\frac{P(S_2)-P(S_1)}{P(S_1)}}\nonumber\\
\le&1+\frac{P_{\max}^{n/2}}{\sqrt{P(S_1)}}\triangleq1+c_3P_{\max}^{n/2}
\end{align}
Similarly,
$
||M^{-\frac{1}{2}}||_2^{-1}\ge1-c_4P_{\max}^{n/2}
$. 
From Lemma \ref{times}, we have
\begin{equation}
(1-c_4P_{\max}^{n/2})\le\frac{\lambda_2(\tilde{P})}{\lambda_2(M^{-\frac{1}{2}}\tilde{P})}
\le(1+c_3P_{\max}^{n/2})
\end{equation}
Since $P$ is a joint distribution matrix, from Theorem \ref{iff}, we know that $\lambda_2(\tilde{P})\le 1$.
Therefore, we have
\begin{align}
(1-c_4P_{\max}^{n/2})(1-c_2P_{\max}^{n/2})&\le\lambda_2(\tilde{P})\le1
\end{align}
When $P_{\max}<1$, corresponding to the non-trivial case,
$\lim_{n\rightarrow\infty} P_{\max}^{n/2}= 0$, 
and using (\ref{lim}), (\ref{mapping}) follows.

The case $P(S_2)< P(S_1)$
can be proved similarly. 
$\qquad\quad\blacksquare$ 

\bibliographystyle{IEEEtran}
\bibliography{IEEEabrv,ref}

\end{document}